\documentclass[12pt]{article}
\usepackage{epsfig}
\usepackage{graphicx}
\usepackage{amsmath}
\usepackage{amstext}
\usepackage{amsfonts}
\usepackage{amssymb}

\def\beq{\begin{equation}}
\def\eeq#1{\label{#1}\end{equation}}
\def\eeqn{\end{equation}}

\def\beqa{\begin{eqnarray}}
\def\eeqa#1{\label{#1}\end{eqnarray}}
\def\eeqan{\end{eqnarray}}

\let\bar=\overbar

\def\Dslash{\not{\hbox{\kern-4pt $D$}}}
\def\dslash{\not{\hbox{\kern-2pt $\del$}}}

\def\msb{{\bar{\ssstyle M \kern -1pt S}}}

\def\Title#1{\begin{center} {\Large {\bf #1} } \end{center}}

\begin{document}

\Title{Charm mixing in the Standard Model:\\ the inclusive approach}

\bigskip\bigskip


\begin{raggedright}  

{\it Markus Bobrowski\index{Bobrowski, M.}\\
Institut f\"ur Theoretische Teilchenphysik\\
Karlsruhe Institute of Technology\\
D-76131 Karlsruhe, GERMANY}
\bigskip\bigskip
\end{raggedright}

\begin{raggedright}
{\footnotesize Proceedings of CKM 2012, the 7th International Workshop on the CKM Unitarity
Triangle,\\ University of Cincinnati, USA, 28 September -- 2 October 2012}
\bigskip\bigskip
\end{raggedright}

Flavour oscillations of neutral kaons and $B$ mesons have been key observations leading to the development of the Standard Model of elementary particle physics. Today, still,  mixing and CP violation play a preeminent role in extracting its fundamental parameters and constraining possible extensions. Oscillations in the neutral $D$ system are experimentally established \cite{DMIX} and can be probed at a precision level at LHCb  \cite{Bediaga:2012py}; they appear in presence of   non-zero mass and width splittings, $\Delta M$ and $\Delta \Gamma$, between the stationary eigenstates.  The  HFAG world averages \cite{Amhis:2012bh} for these quantities are 
\begin{equation} 
\label{eq:xyexp}
\begin{aligned} 
  x &\equiv &\frac{{\Delta M}}{\Gamma } &= \left( {0.63^{+0.19}_{-0.20}}
  \right)\% ,\\[.5ex] 
  y &\equiv &\frac{{\Delta \Gamma }} {{2\Gamma }} &=
  \left( {0.75 \pm 0.12} \right)\% .
\end{aligned}
\end{equation}

On the theoretical side, mixing and weak decays of mesons can be assessed in parton-level perturbation theory using the heavy quark expansion \cite{HQE}, provided there is one heavy constituent quark; the approach was very successful to predict lifetime, mixing rates, and CP violation of neutral and charged $B$ mesons with high accuracy. A 2012 measurement of the $B_s$ lifetime at LHCb, CDF, and D0 \cite{EXP} using angular analysis in $B_s \to \psi\phi$ \cite{Dunietz:2000cr} was in perfect agreement with an earlier Standard Model prediction \cite{LENZNIERSTE}.  The neutral $B_s$ width splitting, measured by LHCb at the $5\sigma$-level  \cite{EXP}, could be predicted  with an accuracy of about 30\% \cite{LENZNIERSTE}, which is an impressive result:  given an energy release as low as $1.4\,\rm{GeV}$ in $B_s \to D_sD_s$, which dominantly generates $\Delta \Gamma$, the heavy-quark expansion had previously been expected to be challenged by violation of quark-hadron duality. 

Compared to the other neutral meson systems $K^0$, $B^0$ and $B_s^0$, mixing in charm is more involved to understand theoretically. It is well known that a straightforward application of the HQE framework fails to predict the mixing rates correctly;  this is often regarded as a sign for the breakdown of the quark-level picture. Indeed, it is not clear whether or not $1/m_c$-suppressed corrections can still be treated pertubatively at the charm threshold. QCD radiative corrections, in addition, will have a larger impact at the lower scale. Still, there are some observations suggesting that the heavy-quark framework should not be that  unreliable in charm: the energy released in the decays dominantly generating $\Delta\Gamma$, for example, is not vastly different for the $D$--$\bar D$ and $B$--$\bar B$ transition. The expansion parameter in the heavy-quark expansion is, roughly speaking, some scale of hadronic dynamics over energy release; the $1.4\,\rm{GeV}$ quoted above for $B_s \to D_sD_s$ compare to $1.6\,\rm{GeV}$ in $D \to \pi\pi $, or $0.9\,\rm{GeV}$ in $D \to KK$. In $\Delta \Gamma (B_s)$, in addition, it turned out that the actual expansion parameter is even smaller than the naive expectation, pointing towards an effective hadronic scale well below $1\,\rm{GeV}$ \cite{LENZNIERSTE}.  The impression is supported when looking at the structure of the perturbative expansion of $\Delta\Gamma(D)$: albeit at a wrong central value, next-to-leading order corrections in both QCD and $1/m_c$ add not more than 50\% correction to the Wilson coefficients \cite{LIFETIMES}. 

\begin{figure}[t]
\centering
\includegraphics[height=.18\textwidth]{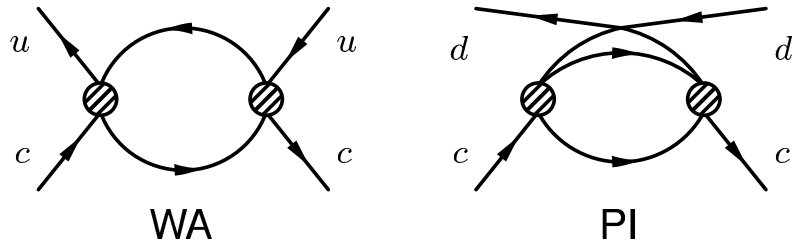}
\caption{$1/m_c$-leading contributions to the decay width $\Gamma\left(D^0\right)$ and $\Gamma\left(D^+\right)$. They contribute to the effective Hamiltonian through operators of dimension six and probe  heavy-quark expansion in the same order as the $D^0$--$\bar D^0$ mixing rate.}
\label{fig:diagrams}
\end{figure}

\begin{figure}[t]
\centering
\includegraphics[width=.45\textwidth]{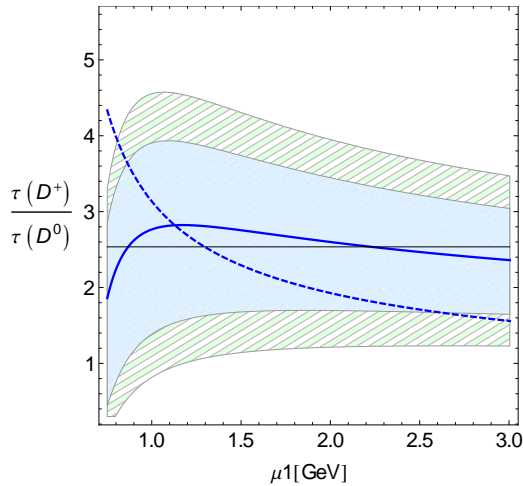}
\caption{Standard Model estimate for $\tau\left({D^+}\right)/\tau\left({D^0}\right)$ at operator dimension six, as a function of the $\Delta C=1$ renormalization scale $\mu_1$ in the $\overline{\rm MS}$ renormalization scheme, at leading (dashed) and next-to-leading order (solid). The error bands correspond to a plausible range of the most relevant hadronic parameters \cite{Bobrowski:2012jf}. The thin black line indicates the PDG lifetime ratio.}
\label{fig:lifetimes}
\end{figure}

I discussed recently \cite{Bobrowski:2012jf} that there is support for the OPE approach also from the charm meson lifetimes. Both $D^0$ mixing and the $D^+$--$D^0$ lifetime difference  are effects beyond the spectator model quark decay, see Fig. \ref{fig:diagrams}. Their leading contribution is from operators of dimension six and tests the heavy-quark expansion in the same order in $1/m_c$. There is one important difference, though: meson-antimeson mixing is an SU(3) breaking effect; zero in the limit of strict SU(3), mixing rates are proportional to powers the light quark masses, which are the spurions of the SU(3) symmetry breaking. The $1/m_c$-leading operator-dimension six therefore predicts the  $D^0$--$\bar D^0$ transition to proceed much slower than what is actually the case. I will argue below that specific operators of higher order in the heavy-quark expansion do not show this feature and turn out to dominate the  series.  The $D^+$--$D^0$ lifetime difference, in converse, does not know this kind of SU(3) suppression already in  dimension six. As a consequence, for the heavy-quark expansion to work out successfully, the latter will have to reproduce the measured values within the first few orders. Indeed, already at leading order in  $1/m_c$ the agreement with the PDG numbers \cite{Beringer:1900zz} is surprisingly good: 
 \begin{equation}\label{lifetimes}
 \begin{aligned}
\frac{\tau\left({D^+}\right)}{ \tau\left({D^0}\right)}_{\rm exp} &= 2.536 \pm 0.019\; , \\[.5ex]
\frac{\tau\left({D^+}\right)}{ \tau\left({D^0}\right)}_{\bar{\rm MS}}  &= 2.8 \pm 1.5^{\;\rm (hadronic)} \;_{-0.7}^{+0.3\;\rm (scale)} \pm 0.2^{\;\rm (exp)} \; . \\[.5ex] 
\end{aligned}
\end{equation}
This estimate includes QCD at next-to-leading order, renormalized in the $\overline{\rm MS}$ scheme. I show the scale dependence in Fig. \ref{fig:lifetimes}. Perturbation theory appears to be reliable above around $1\,\rm{GeV}$: next-to-leading QCD corrections are moderate in size and decrease the scale dependence. The large error in (\ref{lifetimes}) is mainly due to poor knowledge of the relevant operator matrix elements. Better lattice data could reduce this error significantly. I see these results justifying some confidence in the short-distance picture. Note, however, that we should expect the subleading dimension seven to add considerable corrections, which will offset the very good agreement with the measured lifetime ratio. An estimate of this effect will be essential to understand the convergence behaviour of the $1/m_c$ expansion. 

\begin{figure}[t]
\centering
\includegraphics[width=\textwidth]{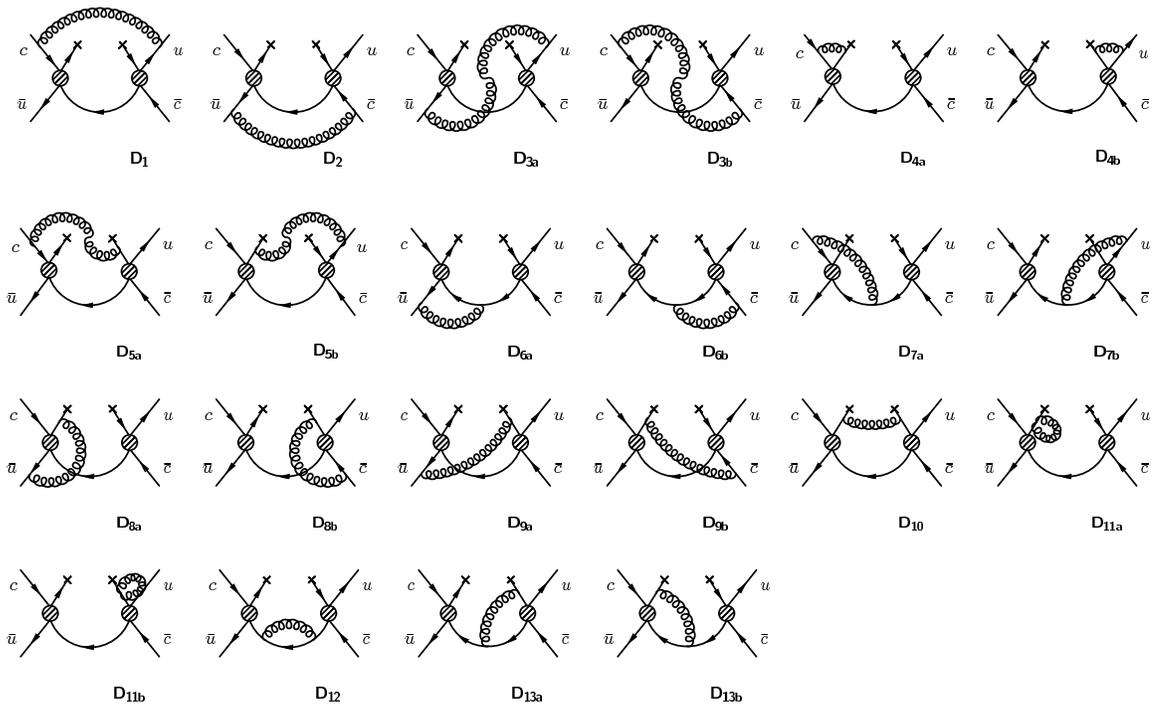}
\caption{Contributions to $\Delta \Gamma$ coupling to the soft hadronic background. The gluon is necessary to generate an absorptive part.}
\label{fig:cdsLandscape}
\end{figure}

In a naive application of the heavy quark expansion to mixing, $D^0$--$\bar D^0$ mixing rates are predicted  far too small. As mentioned earlier, this is due to a suppression by powers of the SU(3) breaking parameter $m_s/m_c$. It has been realized already long time ago \cite{SU3}, that the effect is bypassed once the intermediate state couples to sources of SU(3) breaking other than the quark masses. In fact, SU(3) breaking can be introduced by non-perturbative QCD dynamics: particles with soft momenta in the intermediate state of the $\Delta C=2$ transition become sensitive to the soft QCD background in the hadron and feel the SU(3) breaking in the hadron state. In the operator product expansion this interaction contributes through six-quark operators of dimension nine, see Fig. \ref{fig:cdsLandscape}. The mild suppression by three additional powers of the charm mass is easily overcompensated by the stronger SU(3) breaking, making the effect the dominant contribution to the width splitting $\Delta \Gamma$, without spoiling the convergence of the $1/m_c$ expansion.  To obtain a numerical estimate for the hadronic matrix elements of six-quark operators  we assume vacuum saturation \cite{Shifman:1978bxyw} for the soft sea-quark pair and approximate its hadronic matrix element with the QCD vacuum condensate. 

A calculation of the contributions illustrated in Fig. \ref{fig:cdsLandscape} confirms the expectation expressed above: the correction from operator-dimension nine to a transition with a specific single-flavour intermediate state is in the percent-range. This is reassuring and demonstrates that up to operator-dimenson nine the  corrections point towards a convergence of the heavy-mass expansion. Summing over all intermediate state flavours, the SU(3) breaking in the QCD vacuum protects  operator-dimenson nine from the severe cancellation affecting the $1/m_c$-leading orders, leaving it as the  dominant effect. Numerically, it exceeds the latter by a factor of order ten in the mixing rate, which is found to be 
\begin{equation}
  y = \left( {8 \pm 9} \right) \cdot 10^{ - 6} . 
\end{equation}
Albeit still far below the observation, the result demonstrates that SU(3) breaking from the hadron state is an important effect. A similar contribution appears once more in operator dimension twelve, when all intermediate state quarks in the $D$--$\bar D$ transition are soft, and may help to move the short-distance expectation closer towards the measurement.

\section*{Acknowledgements}
I would like to thank the organizers of \textit{CKM 2012} for the invitation and a pleasant workshop. I gratefully acknowledge the support by grants of the German National Academic Foundation, the State of Bavaria, the Elite Network of Bavaria, and the German Academic Exchange Service. My work has further been supported by the DFG Research Unit SFB/TR9.

\end{document}